\RequirePackage{fix-cm}

\documentclass[twocolumn]{svjour3}          
\smartqed  
\usepackage{graphicx}
\journalname{Visual Computer journal} 
\begin{document}

\title{Scenior: An Immersive Visual Scripting system based on VR Software Design Patterns for Experiential Training}
\author{Paul Zikas \and George Papagiannakis \and Nick Lydatakis \and Steve Kateros \and Stavroula Ntoa \and Ilia Adami \and Constantine Stephanidis
}


\institute{Paul Zikas \at
              ORamaVR \\
              \email{paul@oramavr.com}           
           \and
           George Papagiannakis \at
              ORamaVR, Institute of Computer Science, Foundation for Research and Technology-Hellas, University of Crete, Department of Computer Science \\
              \email{george.papagiannakis@oramavr.com} 
                         \and
           Nick Lydatakis \at
              ORamaVR \\
              \email{nick@oramavr.com} 
                         \and
           Steve Kateros \at
              ORamaVR \\
              \email{steve@oramavr.com} 
                         \and
           Stavroula Ntoa \at
              Institute of Computer Science, Foundation for Research and Technology-Hellas  \\
              \email{stant@ics.forth.gr} 
                         \and
           Ilia Adami \at
              Institute of Computer Science, Foundation for Research and Technology-Hellas  \\
              \email{iadami@ics.forth.gr} 
                         \and
           Constantine Stephanidis \at
              Institute of Computer Science, Foundation for Research and Technology-Hellas, University of Crete, Department of Computer Science  \\
              \email{cs@ics.forth.gr} 
}
\date{Received: 29/02/2020 / Accepted: 29/04/2020}

\maketitle

\begin{abstract}
Virtual reality (VR) has re-emerged as a low-cost, highly accessible consumer product, and training on simulators is rapidly becoming standard in many industrial sectors. However, the available systems are either focusing on gaming context, featuring limited capabilities or they support only content creation of virtual environments without any rapid prototyping and modification. In this project, we propose a code-free, visual scripting platform to replicate gamified training scenarios through rapid prototyping and VR software design patterns. We implemented and compared two authoring tools: a) visual scripting and b) VR editor for the rapid reconstruction of VR training scenarios. Our visual scripting module is capable to generate training applications utilizing a node-based scripting system whereas the VR editor gives user/developer the ability to customize and populate new VR training scenarios directly from the virtual environment. We also introduce action prototypes, a new software design pattern suitable to replicate behavioral  tasks for VR experiences. In addition, we present the training scenegraph architecture as the main model to represent training scenarios on a modular, dynamic and highly adaptive acyclic graph based on a structured educational curriculum. Finally, a user-based evaluation of the proposed solution indicated that users - regardless of their programming expertise - can effectively use the tools to create and modify training scenarios in VR.
\end{abstract}

\section{Introduction}
Virtual reality has advanced rapidly, offering highly interactive experiences, arousing interest in both the academic and the industrial community. VR is characterized by highly immersive and interactive digital environments where user experiences another dimension of possibilities. As already known from conducted trials \cite{HOOPER2019}, \cite{SkillTran}, the training capabilities of VR simulations offer skill transfer from the VR to real-life proposing an effective tool to fit in modern curricula. From pilots to surgeons, VR has a strong impact on training due to embodied cognition, psychomotor capabilities (dexterous use of hands) and high retention level \cite{greenleafVR}.

Authoring tools encapsulate key software functionalities and features for content creation. The software architecture of such system empowers programmers with the necessary tools for content creation. However, existing platforms do not sufficiently propose a complete methodology to reconstruct a training scenario in virtual environments. Training simulations are often implemented in modern game engines using native tools without any customization or specially designed features for generating of immersive scenarios rapidly. In addition, there are are no prototyped software patterns specially formulated for VR experiences, leading to complex implementations and lack of code reusability.

Previously we have proven that our VR training platform \cite{MAGES2018} makes medical training more efficient. In a revolutionary clinical study \cite{HOOPER2019} in cooperation with New York University that established - for the first time in the medical bibliography - skill transfer and skill generalization from VR to the real Operating Room in a quantifiable, measurable ROI.

In this project, we propose a visual scripting system capable to generate VR training scenarios following a modular Rapid Prototyping architecture. Our initial goal was to define how to construct complex training pipelines from elementary behaviors  derived from prototyped and reusable software building blocks. Inspired from game programming patterns, we implemented new software design patterns named \emph{Actions} for VR experiences to support a variety of commonly used interactions and procedures within training scenarios offering great flexibility in the development of immersive VR metaphors. We designed our solution as a collection of authoring tools combining a visual scripting system and an embedded VR editor forming a bridge from product conceptualization to product realization and development in a reasonably fast manner without the fuss of complex programming and fixtures. Our goals and design decisions were the following:

\begin{itemize}
\item \textbf{Educational pipeline:} We are interested in representing an educational process into an efficient data structure, for simple creation, easy maintenance and fast traversal.
\item \textbf{Modular Architecture:} To support a wide variety of interactions and different behaviors within the virtual environment we want our system to integrate a modular architecture of different components linked into a common structure. 
\item \textbf{Code-free SDK:} Our intentions were to develop a platform where users can create VR training scenarios without advanced programming knowledge. We also want to study techniques for the visual creation of VR experiences.
\item \textbf{Rapid Prototyping:} We are interested in making reusable prototyped modules to implement more complex interactive behaviors derived from elementary blocks. Our goal is to define basic structural elements capable to visualize simple behaviors but when combined recreate complex scenarios.
\item \textbf{VR Software Design Patterns:} What are the benefits of gamified software design patterns in VR applications? We aim to support a large number of interactive behaviors in VR applications to promote new software patterns specially formulated to speed up content creation in VR. 
\end{itemize}

This paper is organized as follows. In Section 2 we present the state of the art in training simulations and similar authoring tools. In Section 3 we introduce our solution with a brief description of our software modules. Section 4 presents the training scenegraph architecture. Section 5 describes the Action Prototypes and our rapid prototyping solution. Section 6 presents the visual scripting tool and Section 7 the VR editor. In Section 8 we discuss our results from the evaluation process. Section 9 concludes and defines the future work.

\begin{figure*}
  \includegraphics[width=\textwidth]{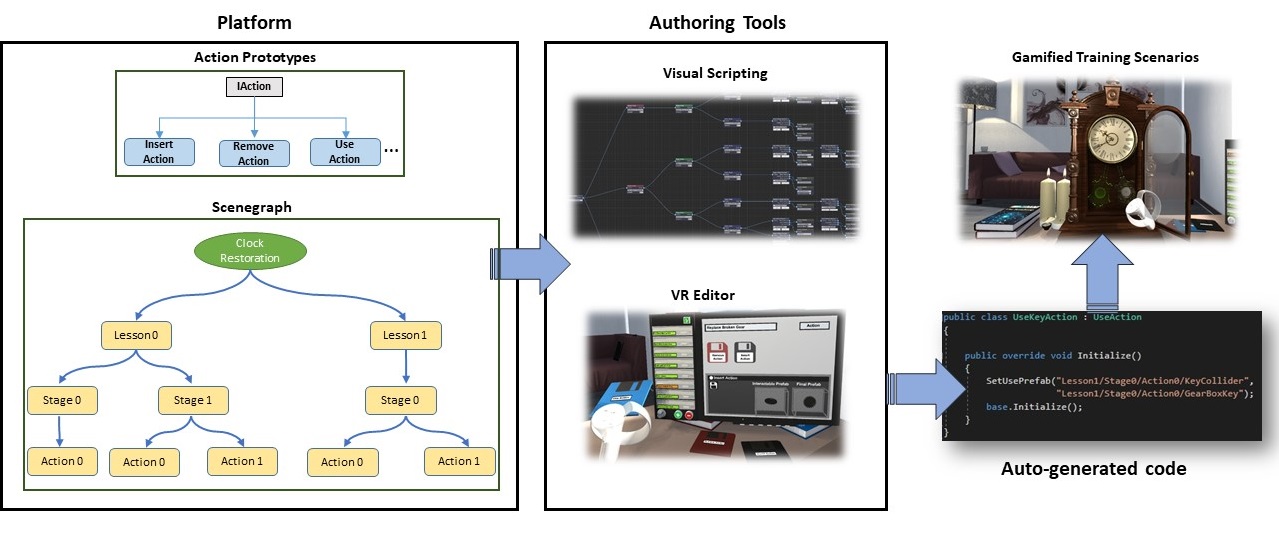}
  \caption{The architectural diagram of our system. The platform consists of a training scenegraph along with the action prototypes. In a higher hierarchy, the authoring tools (visual scripting and VR editor) are facilitating tools to generate interactive behaviors in the virtual environment. Finally, the training scenarios are implemented from the auto-generated code.}
  \label{fig:VSEditor}
\end{figure*}

\section{Related Work}
In this section, we present the state of the art in VR training, its impact on education and similar authoring platforms.

\subsection{The impact of VR in training and education}
The engagement of education with novel technological solutions provide new opportunities to increase collaboration and interaction through participants, making the learning process more active, effective, meaningful, and motivating \cite{Alsumait2013}. Collaborative VR applications for learning \cite{Greenwald2017TechnologyAA}, studies for the impact of VR in exposure treatment \cite{exposureVR} as well as surveys for human social interaction \cite{WhyVR} have shown the potential of VR as a training tool. The cognitive aspect of VR learning is already known from conducted trials \cite{GanierCognitive}. Recent examples are featuring the learning capabilities of VR in surgical simulations \cite{Papagiannakis2018} with remarkable results.

Focusing on the educational factor, the use of VR for knowledge transfer and e-learning is now extended as the R\&D grows around entire VR environments where the learning takes place \cite{Teresa}. Virtual Reality rapidly increases its potential and influence on e-learning applications \cite{elearn} by taking advantage of two basic principles: a) the embodiment \cite{Slater2017} and b) the increased knowledge retention \cite{Retention} with immersive environments capable to present a realistic scenario as it is, as it would be in real-life.

\subsection{Authoring tools for content creation}
The main concept behind authoring tools is to develop frameworks capable to generate content with minimal changes to speed up content creation while improving product maintenance.

BricklAyeR \cite{BricklAyeR} proposes a collaborative platform designed for users with limited programming skills that allows the creation of Intelligent Environments through a building-block interface. Another interesting project is ARTIST \cite{ARTIST}, a platform that provides tools for real-time interaction between human and non-human characters to generate reusable, low cost and optimized MR experiences. Its aim is to develop a code-free system for the deployment and implementation of MR content while using semantically data from heterogeneous resources.

Another authoring tool, ExProtoVAR \cite{ExProtoVAR} generates interactive experiences in AR featuring development tools specially designed for non-programmers, without necessarily  a technical background with AR interfaces. In the field of interactive storytelling, StoryTec \cite{StoryTec} platform facilitates an authoring tool to generate and represent storytelling-based scenarios in various domains (serious games, e-learning and training simulations). The platform aims to standardize the content creation of storytelling experiences following a descriptive format. 

\subsection{Visual Programming}
Visual programming is getting more publicity as more platforms and tools are emerging. We can separate them into two categories according to their visual appearance and basic functionalities: a) block-based and b) node-based scripting languages

Block-based visual languages consist of modular blocks that represent fundamental programming utilities. OpenBlocks \cite{Roque2008} proposes an extendable framework that enables application developers to build a custom block programming system by specifying a single XML file. Google's online visual scripting platform Blocky \cite{Pasternak2017} uses interlocking, graphical blocks to represent code concepts like variables, logical expressions, loops, and other basic programming patterns to export blocks to many programming languages like JavaScript, Python, PHP and Lua. Another interesting approach is the Scratch \cite{Maloney2010} visual programming language which primarily targets ages 8 to 16 offering an authoring tool to support self-directed learning through tinkering and collaboration with peers.

On the other hand, node-based visual languages, represent structures and data flow using logical nodes to reflect a visual overview of data flow. GRaIL \cite{Ellis1969} was one of the first systems that featured a visual scripting method for the creation of computer instructions based on cognitive visual patterns. It was used to make sophisticated programs that can be compiled and run at full speed, or stepped through with a debugging interpreter. More recently, \cite{Kensek2015} published three case studies on visual programming for building information modeling (BIM) utilizing Dynamo, a graphical programming framework. In addition, Unity3D game engine has recently announced at their 2020 roadmap \cite{UnityRoadmap} an embedded node-based editor and a visual scripting system that will launch with their next update.

\subsection{Editing directly from the VR environment}
The development of authoring tools in virtual reality systems led to the integration of sophisticated functionalities. One of them is the implementation of immersive VR editors for the reconstruction of digital worlds directly from within the virtual environment.

In SIGGRAPH 2017, Unity technologies presented EditorVR \cite{EditorVR}, an experimental scene editor that encapsulates all the Unity's features within the virtual environment giving developers the ability to create a 3D scene while wearing the VR headset. EditorVR supports features for initially laying out a scene in VR, making adjustments to components and building custom tools.

Except from game engines, model editors are also emerging into immersive VR model editing.  MARUI \cite{MARUI} is a plugin for Autodesk Maya that lets designers perform modeling and animation tasks within the virtual environment. Another noticeable project is RiftSketch \cite{Elliott2015}, a live coding VR environment, which allows the development and design of 3D scenes within the virtual space. RiftSketch proposes a hybrid XR system utilizing an external RGB camera and a leap motion sensor to record live footage from the programmer's hands while coding and project this image into the virtual environment.

The available VR editors feature scene management capabilities with intuitive ways to build a scene directly from within the virtual environment. However, there are no available authoring tools to offer a complete system for developing a behavioral VR experience including both the design and the programming aspect.

The state of the art shows that VR platforms do not provide sufficient tools to generate training simulations nor a complete methodology for representing an educational process in VR.

\section{Our Solution}
The main goal of this project is to implement and compare three different authoring mechanics a) prototyped scripting, b) visual scripting and c) VR editor for rapid reconstruction of VR training scenarios based on our newly defined VR software design patterns. In more detail, the proposed system facilitates a VR playground to recreate training scenarios using the developed tools and functionalities. From the developer's perspective, this system constitutes a Software Development Kit (SDK) to generate VR content, which follows a well-structured educational pipeline. After coding the training scenario, users can experience the exported simulation.

For rapid operation adaptation to variations, we implemented a schematic representation of VR experiences to replicate training scenarios in a directed acyclic graph. By prototyping commonly used interaction patterns we managed to create a customizable platform able to generate new content with minimal changes. Inspired from game programming patterns, we implemented new design patterns for VR experiences to support a variety of commonly used interactions and procedures within training scenarios offering great flexibility in the development of VR metaphors. We built our system as a plugin for Unity3D engine for effective setup and distribution.

We introduce the following contributions: 
\begin{itemize}
\item \textbf{Training Scenegraph:} We developed a dynamic, modular tree data structure to represent the training scenario following a well defined educational curriculum. A training scenegraph tree stores data regarding the tasks where the trainee is asked to accomplish, dismantling the educational pipeline into simplified elements and focusing on one step at the time.
\item \textbf{Action Prototypes:} We designed reusable prototypes based on VR software design patterns to transfer behaviors from the real to the virtual world. Action prototypes populate the training scenegraph with interactive tasks for the user to accomplish. They introduce a novel methodology specially formulated for the development of interactive VR content.
\item \textbf{Visual Scripting:} We integrated a Visual Scripting system as an authoring tool to export training scenarios from a node-based, coding-free user interface.
\item \textbf{VR Editor:} We embedded a run-time VR Editor within the training scenarios to give user the ability to customize and create new scenarios directly from within the virtual environment. 
\item \textbf{Pilot applications:} Utilizing the proposed system we generated two pilot training scenarios: a) a REBOA (Resuscitative Endovascular Balloon Occlusion of the Aorta) training scenario and b) an antique clock restoration.
\end{itemize}

In the following sections, we present the software modules and key functionalities of our system.

\section{The Training Scenegraph}

To achieve a goal, whether it is the restoration of a statue, the repair of an engine's gearbox or a surgical procedure the trainee needs to follow a list of tasks. We are referring to those tasks as \emph{Actions}. 

A simple visualization of a training scenario containing Actions would be to link them in a single line one after another. However, in complex training scenarios, a sequential representation would not be very convenient due to the absence of classification and hierarchical visual representation. For this reason, we implemented the training scenegraph architecture. Training scenegraph is a tree data structure representing the tasks/Actions of a training scenario. The root of the tree holds the structure, on the first depth we initialize the \textit{Lesson} nodes, then the \textit{Stage} nodes and finally at leaf level the \textit{Action} nodes.

\begin{figure}[!htbp]
  \includegraphics[width=\linewidth]{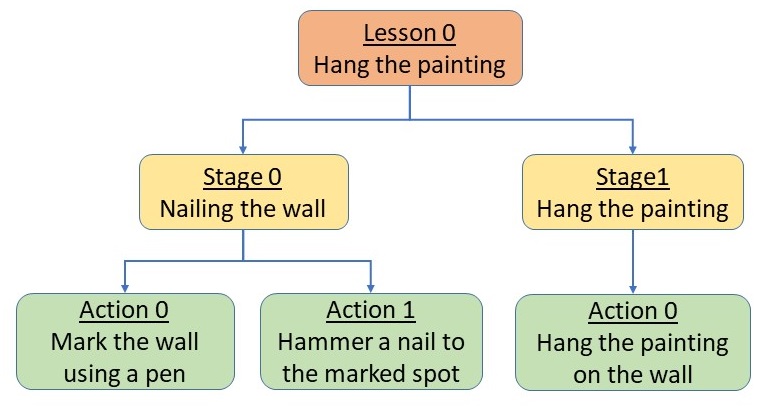}
  \caption{An example of a training scenegraph tree representing the simple scenario of hanging a paint on the wall.}
\end{figure}

One of the main principles of this project was to modify the training scenegraph and Actions using three different editors (scripting, visual scripting and the VR editor). To achieve this, the scenegraph data is stored to an xml file offering extended functionalities, editing abilities and easy maintenance of the scenegraph structure even for complex training scenarios.

\section{Action Prototypes}
In this section, we analyze how we implemented new VR software design patterns thought rapid prototyping.

\subsection{The IAction Interface}

The Action object reflects a flexible structural module, capable to generate complex behaviors from basic elements. This also reflects the concept idea behind the training scenegraph; provide developers with fundamental elements and tools to implement scenarios from basic principles. Each Action is described by a script containing its behavior in means of physical actions in the virtual environment. Technically, each Action script implements the IAction interface, which defines the basic rules every Action should follow, ensuring that all Actions will have the same methods and structure. Below we present the components of IAction interface.

\begin{itemize}
\item  \textbf{Initialize:} This method is responsible to instantiate all the necessary 3D objects for the Action to operate.
\item  \textbf{Perform:} This method completes the current Action and deletes unused assets before the next Action starts.
\item  \textbf{Undo:} This method resets an Action including the deletion of instantiated 3D assets and the necessary procedures to set the previous Actions.
\item   \textbf{Clear:} Clears the scene from initialized objects and references from the Action.
\end{itemize}

Designing a shared interface among the structural elements of a system is the first step to prototype commonly used components. This methodology is both beneficial for the user and the developer: a) users are introduced with interactive patterns that are familiar with, avoiding complex behaviors while b) developers are following the same implementation patterns.

\subsection{From Actions to VR Design Patterns}
To make our system more efficient we have to limit the capabilities of the Action entity targeting simple but commonly used tasks in training. Modeling those behaviors, we will generate a pool of generic behavioral patterns suitable for VR applications.

The implementation of Action prototypes was highly inspired by Game Programming Patterns \cite{Nystrom2014} as an alternative paradigm for VR experiences. The immersion of virtual environments causes the implementation of programming patterns to fit into a more interactive way of thinking. For this reason, the software patterns developed in this project designed to match the needs for interactivity, embodied cognition and physicality on VR experiences. For this reason, we implemented the following Action Prototypes: 

\begin{itemize}
\item  \textbf{Insert Action:} is referring to the insertion of an object to a predefined position. Technically, to implement an Insert Action the developer needs to set the initial and the final position of an object, then the task for the user would be to take this particular object and place it in the correct position paying respect to its orientation.
\item \textbf{Remove Action:} describes a step in which the user has to remove an object using his hands. To implement a Remove Action the developer needs to define the position where the object will be instantiated, for user to reach and remove it.
\item \textbf{Use Action:} refers to a step where the user needs to interact with an object over a predefined area for a period of time.

\begin{figure}
  \includegraphics[width=\linewidth]{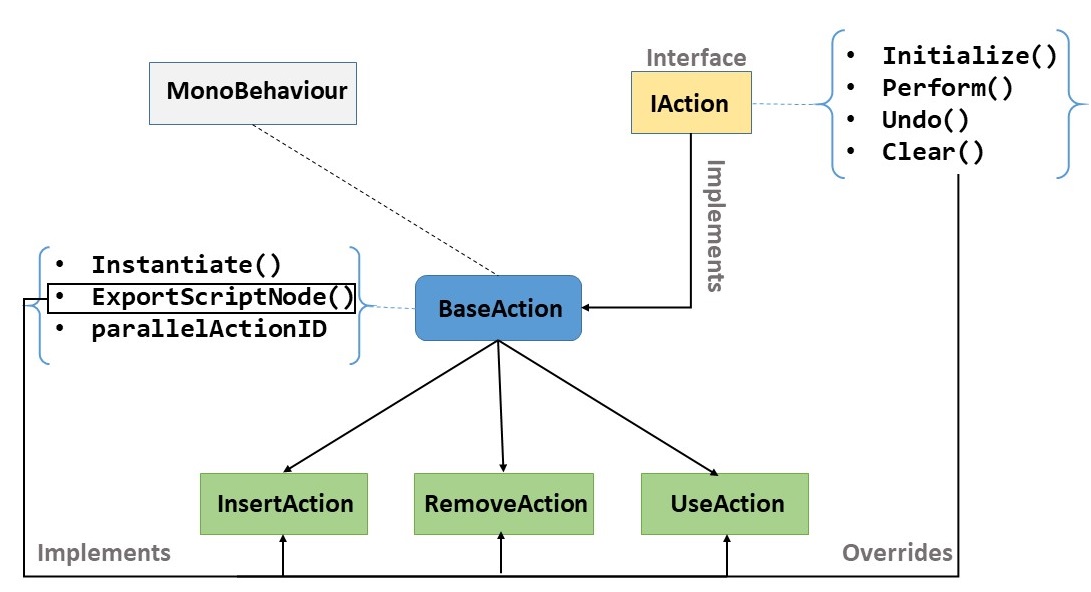}
  \caption{Action Prototypes Architecture diagram.}
  \label{fig:ActionPrototypesDiagram}
\end{figure}

Figure~\ref{fig:ActionPrototypesDiagram}  illustrates an architectural diagram of Action Prototypes to visualize their dependencies.
\end{itemize}

Action Prototypes constitute a powerful software pattern to implement interactive tasks in VR experiences. Unitizing Action Prototypes, developers can replicate custom behaviors with a few lines of code taking advantage of their abstraction and reusability. New Action Prototypes can be easily implemented due to their abstraction and the use of IAction interface.  

\begin{figure}[!htbp]
\begin{center}
  \includegraphics[width=\linewidth]{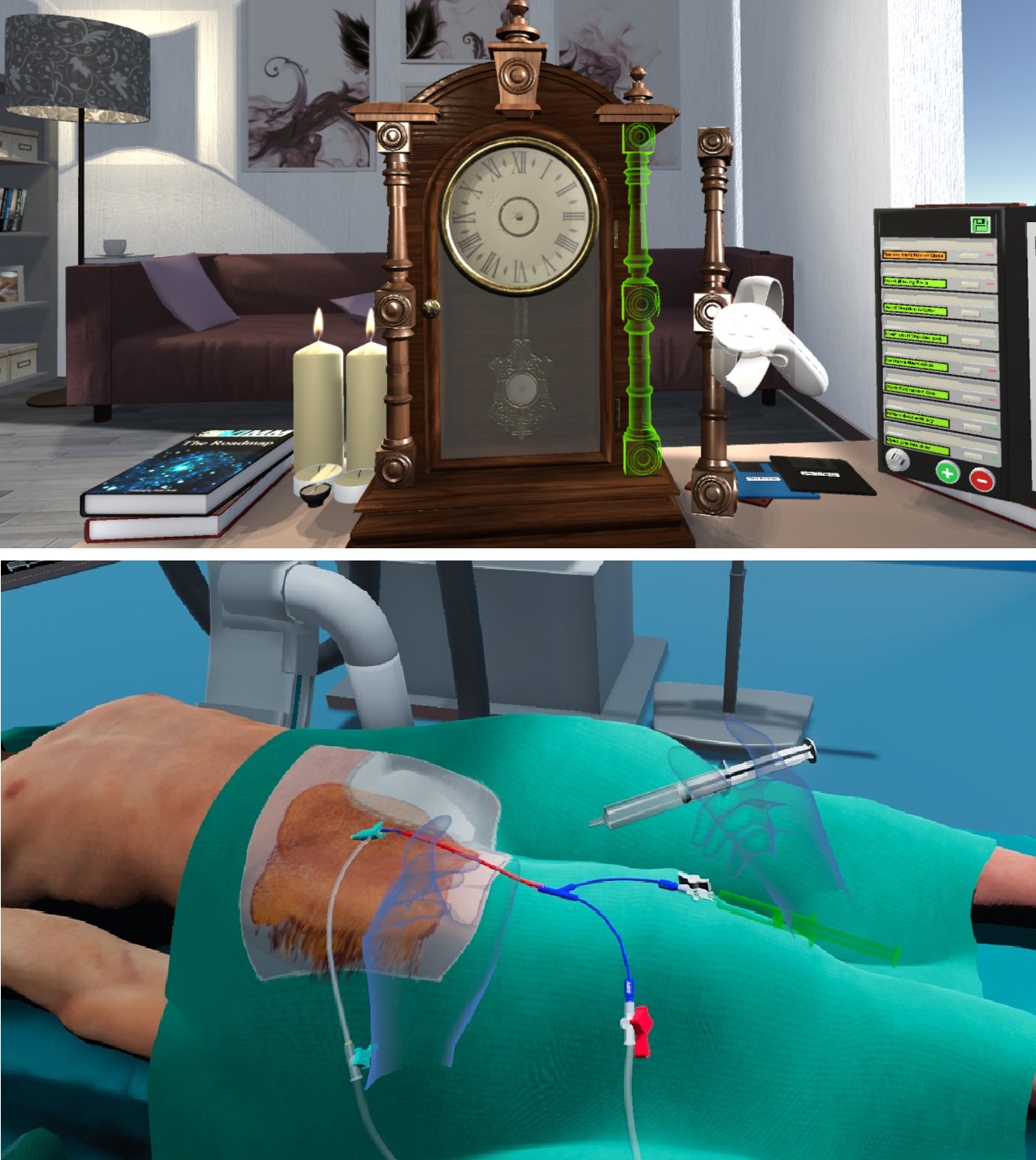}
  \caption{Top: Insert Action from the clock's maintenance use case. Bottom: Insert Action from the REBOA use case. The green holograms indicate the correct position of 3D objects}
  \end{center}
\end{figure}

\subsection{Alternative Paths}

The Action prototypes propose a new design pattern for VR experiences, a modular building block to develop applications in combination with the training scenegraph. However, the proposed training scenegraph architecture generates VR experiences following a "static" pipeline of Actions where user needs to complete a predefined list of tasks. In order to transform the training scenegraph from a static tree into a dynamic graph, we introduced Alternative Paths. 

A training scenario can lead to multiple paths according to the user's actions and decisions, scenegraph adapts. In addition, certain actions or even wrong estimations and technical errors may deviate the original training scenario from its normal path causing the training scenegraph to modify itself accordingly. Except for backtracking after wrong estimations and errors, the Alternative Path mechanic is also used in situations where the trainee needs to make a particular decision over a dilemma.

From a technical perspective, Actions are able to trigger alternative path events. Those events will be advanced to training scenegraph informing about the necessary follow-up actions. The scenegraph tree will update its form accordingly by pruning or adding new nodes to its structure to adapt to the new circumstances. This event will trigger a transformation of the training scenario forcing the user to make additional or different steps due to this differentiation. 

\subsection{The Uncanny Valley of Interactivity and VR UX}

\begin{figure}[!htbp]
  \includegraphics[width=\linewidth]{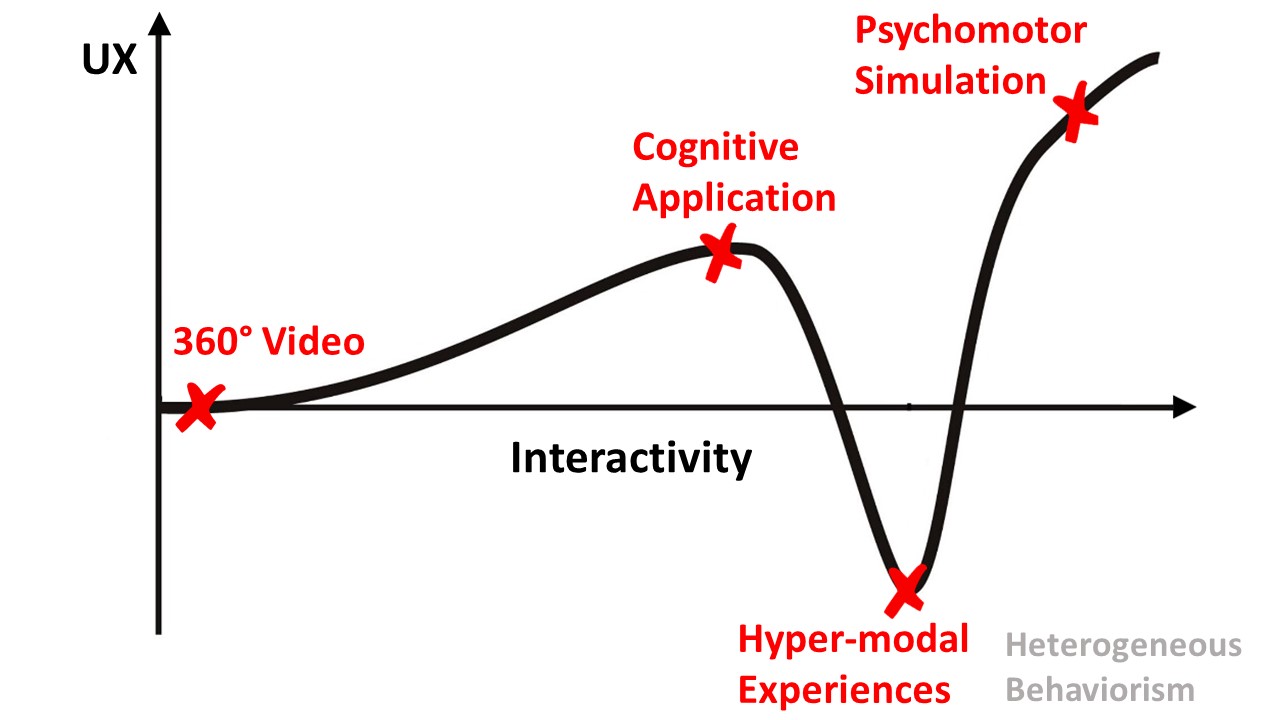}
  \caption{The uncanny valley of interactivity. Correlation between User experience (UX) and interaction in VR.}
  \label{fig:valley}
\end{figure}

After experimenting with various design patterns and interaction techniques for VR, an interesting pattern appeared regarding the correlation of user experience and the interactivity of the VR application (figure~\ref{fig:valley}). An immersive experience relies significantly on the implemented interactive capabilities that form the user experience. As a result, to make an application more attractive in means of UX a more advanced interactive system is needed. However, as we implement more complex interaction mechanics there is a point in timeline where the UX drops dramatically. At this point, the application is too advanced and complex for the user to understand and perform the tasks with ease. We characterize this feature as heterogeneous behaviorism meaning that user's actions do not follow a deterministic pattern resulting in the inability to complete the implemented Actions due to their incomprehensible complexity.

In contrast, applications with limited interactivity follow a linear increase in their user experience. From applications where users are only observers (360VR videos) to cognitive applications, the interaction curve is linear and VR experiences easy to understand. To overcome the effect mentioned before, applications need to drastically enhance their interactivity capabilities and offer users a more intuitive VR environment to understand how they are supposed to act in the virtual world. Overpassing the valley of interactivity, applications are evolving rapidly to follow a psychomotor methodology integrating embodied cognition for better UX. 

\begin{figure*}
  \includegraphics[width=\textwidth]{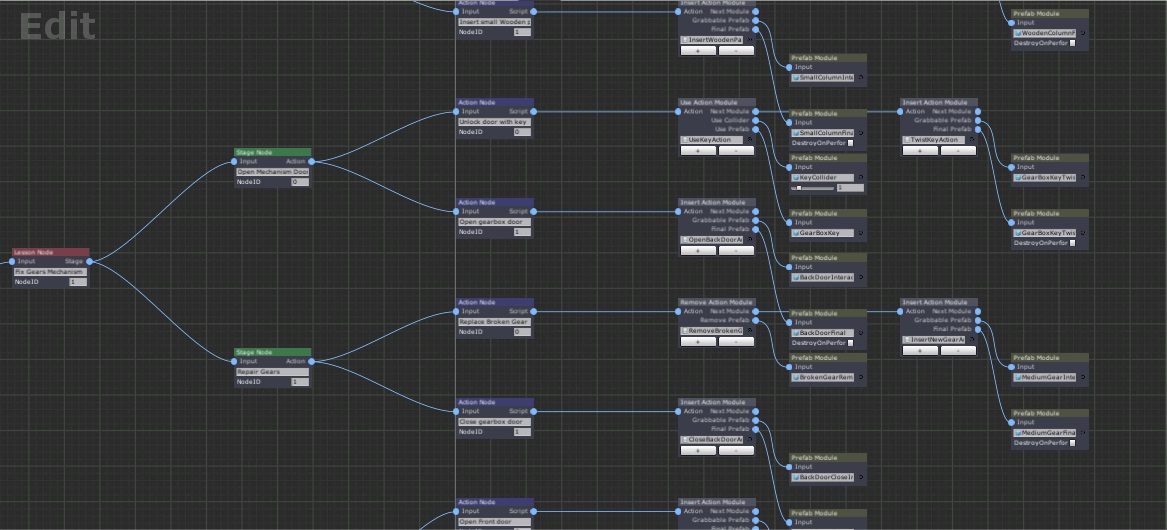}
  \caption{A training scenario visualized from the Visual Scripting Editor featuring from right to left: Lessons (red), Stages (green), Actions (blue), Action Scripts (gray) and Prefab nodes (brown). Prefabs are representing 3D objects in the virtual environment. For example an Insert Action contains two prefabs: the interactable item and its final position. }
  \label{fig:VSEditorBig}
\end{figure*}

\section{Visual Scripting}

The training scenegraph model is capable to generate applications from reusable fundamental elements (Actions) supporting basic insert, remove and use behaviors in VR. However, what is the next step? What can be done to enhance the development process and speed up content creation? The complexity of scenegraph xml may cause difficulties visualizing the scenegraph nodes, especially for extended training scenarios. Another point is the programming skills required do develop such experiences. Using the proposed architecture could be challenging for inexperienced programmers 

To eliminate the mentioned difficulties, we introduce visual scripting as an authoring tool to manage, maintain and develop VR experiences. Visual scripting encapsulates all the functionalities from the base model offering high visualization capabilities, which are very effective especially on extended projects. 

\subsection{The visual scripting metaphor}

The development of a visual scripting system as an assistive tool aimed to visualize the VR training scenario in a convenient way, if possible fit everything into one window. The simplicity of this tool was carefully measured to provide tool used also from non-programmers. From the beginning of the project, one of the main design principles was to strategically abstract the software building blocks into basic elements. The main idea behind this abstraction was the improvement of the visual scripting and VR editor tools since fundamental elements construct a better visual representation than complex ones. To render the visual nodes we exploited Unity Node Editor Base (UNEB), an open-source framework, which provides basic node rendering and management functionalities.

Moving into the visual scripting metaphor, the training scenegraph data structure forms a dynamic tree, visualizing the scenario into a node-based editor with nodes linked together forming logical segments. To construct visual nodes, our system retrieves data from the Action scripts through reflection and run-time compilation. An example of a complete diagram representing a training scenario is illustrated in figure~\ref{fig:VSEditorBig}. Developers can utilize visual scripting to generate training scenarios through interactive UIs. In this way, the content creation is transformed into a coding-free process, encapsulating the system principles into equivalent visual metaphors giving users the ability to generate VR training content without high demand in software background.

\subsection{Dynamic code generation}

Visual scripting generates run-time simple Action scripts utilizing the information provided from the visual input. After completing the visual construction of an Action the next step is to generate the Action script to save the implemented behavior in a C\# code script.

To write C\# code run-time, we used CodeDOM \cite{CodeDOM}, a build-in tool for .NET Framework that enables run-time code generation and compilation. The abstraction of Action prototypes offers an elegant implementation to generate each script using a single virtual method. To finalize the Action script, except the Action Type (Insert, Remove or Use) we also need the interactive behavior. Action prototypes retrieve this information directly from the visual scripting editor through the linked nodes relative to the Action module.

\subsection{Expanding auto-generated scripts}

Visual scripting generates a basic Action script containing the Initialize method, the minimum requirement for an Action to run properly. However, there are cases where developers need to implement significantly complex Action behaviors to enhance use experience with additional information and features for a better experience.

Prototyped Actions were developed using a particular software architecture capable to provide the fundamental facilities but also customize Actions according to the developer's preferences. The Perform method can be overridden directly from the Action script to extend the Action's capabilities. The same principle is applicable to all the other virtual methods defined in the IAction interface (Undo, Initialize, etc.). For additional modifications, the best practice is to edit directly the generated script and override the declared IAction methods. In this way, we maintain simple scripts but also provide custom implementations upon request to fit the training scenario.

\section{VR Editor}

The visual scripting system enhanced the usability and effectiveness of the scenegraph system to generate gamified training scenarios through a coding-free platform. The impact on content creation was very strong due to the additional tools and features that introduced. However, visual scripting lacks on one specific and rather important feature: the ability to design on-the-go behaviors and scenarios directly within the virtual environment. This feature will improve design capabilities while offering an intuitive way to modify applications directly from the virtual environment. 

The implementation of VR editor was designed as an authoring tool on top of the training scenegraph architecture, utilizing the developed features of our system. This interactive tool reduces the time needed to produce training scenarios due to the rapid in-game generation of training scenarios. In addition, certain interactive behaviors are better designed directly from VR instead of a window due to the 3D perspective of the medium.

\subsection{The VR metaphor}

\begin{figure}
\begin{center}
  \includegraphics[width=\linewidth]{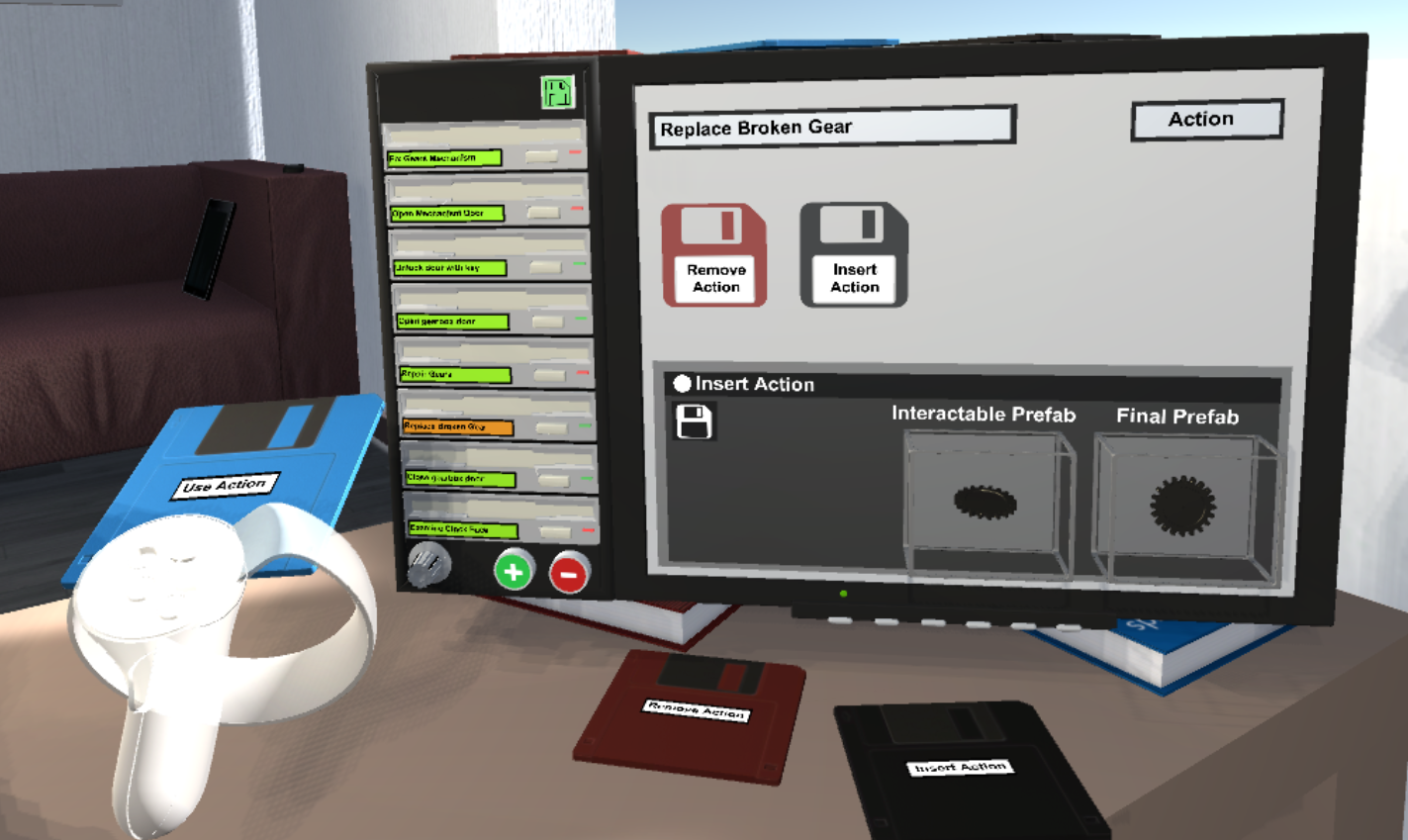}
  \caption{Interacting with the VR editor. User holds a Use Action preparing to generate the Action behavior.}
  \label{fig:VREditorDesign}
  \end{center}
\end{figure}

The main concept behind the implementation of our VR editor focuses on an interactive system with floppy disks and a personal computer. Figure~\ref{fig:VREditorDesign}  illustrates the design of our VR editor along with its various components and floppy disks. The training scenegraph nodes are represented by floppy drives on the left side of the screen. Action scripts are initialized as floppy disks, each one holds the script behavior that defines the 3D objects relative to the Action. There are three types of floppy disk separated with unique coloration; blue disks represent Use Actions, red disk the Remove Actions and black disks the Insert Actions.

The right panel contains the properties of the selected drive. On the top side, we set the name of the scenegraph node along with its type (Lesson, Stage or Action). To implement a new script for an action, the user needs to take a floppy disk and insert it in the corresponding floppy drive.

\subsection{Generate Actions and parametrization on-the-go}

The functionality with the higher impact on the VR editor is by far the ability to modify and parametrize Actions on-the-go. This was also the main reason that led us to implement the VR editor as an additional authoring tool within the virtual environment, to support coding-free development and give the user the ability to modify or even generate new behaviors while playing the training scenario.

Users can customize the scenegraph through VR editor by adding or deleting Scenegraph nodes to match their needs. This functionality has a serious impact on users that want to parametrize existing VR training scenarios or create their own without having any programming knowledge. We provide this ability via an interactive UI on the VR editor with physical buttons and knobs where users can modify and save the training scenegraph.

The next step is the script generation from the VR editor. To generate a new Action, users need to insert a floppy disk into the drive representing the Action script (Insert, Remove or Use). The system will register the insertion of floppy disks and an empty Action script will appear on VR editor screen ready for modification. In a similar way, ejecting a floppy disk from the drive detaches the script from the Action.

With VR editor, users are no longer just observers, they can modify the training scenarios on-the-go, implement new ideas and fix wrong Action behaviors without specialized programming knowledge.

\section{Evaluation}
To examine the overall experience of using our system, we conducted a user-based evaluation with 18 users \cite{EvalNIELSEN}. The main research questions were the following:

\begin{itemize}
\item  For the VR training application: what is the overall perceived quality of the VR training environment and perceived educational value?
\item  For the Visual Scripting tool: can users successfully complete basic programming tasks and how do they rate the overall experience?
\item  For the VR Editor tool: can users successfully complete basic adjustments to an existing training scenario and how do they rate the overall experience?
\end{itemize}

\subsection{Methodology and participants}
The experiment was divided into three separate sessions, one for each tool. In the first session, the participants were asked to restore an antique clock following the instructions given by the VR training application (Clock repair scenario). In the second session, the participants were shown the capabilities and functionalities of the Visual Scripting tool. Then, they were asked to use the tool to generate code for a "Use" action (\emph{'Use the sponge to wipe dirty spot on the clock’}) and a “Remove” action (\emph{‘Remove seal from two-sided gear'}). Finally, in the third session, the participants were asked to complete two tasks to adjust the clock restoration training scenario directly from the VR environment using the VR Editor tool.

A 10-point Likert Scale questionnaire was given at the end of each session to rate the parameters identified in the research questions. For the ‘educational’ value of the VR training application, participants were also asked to indicate which steps they retained regarding the restoration process. In addition, metrics such as the number of help requests and time on task were recorded for further analysis of the results. Finally, at the end of the experiment, a semi-structured interview was conducted to capture participants’ general impression of the whole system.

Eighteen people participated in our experiment, 11 males and 7 females. All users were in the 25-35 age range. They were selected based on the level of expertise in using VR applications and level of expertise in Software Development (SD), ensuring an equal number of expert and non-expert participants in each one of the two categories. \cite{EvalMeasuringUX}, \cite{EvalNIELSEN} 

\subsection{Results}
\subsubsection{First session: VR Training application}
The perceived quality of the VR experience was on average highly rated by all participants (8.6/10). The average rating of VR experts was somewhat lower than that of non-experts, but the difference was not statistically significant, as revealed by a paired t-test analysis (t(8)=-1.83, p=0.1). 

Equally high was the overall average rating score the application received in terms of the perceived educational value (8.78/10). Small differences were exhibited between VR experts and non-experts, however, no statistical significance was identified (t(8)=-0.45, p=0.66).

The participants also scored rather high in the exercise where they indicated which steps they could recall from the restoration process (9.26/10).  A comparison of the achieved score between the two groups did not indicate a statistically important difference (t(8)=-1.08, p=0.31). 

\begin{table}[!http]
\begin{center}
\begin{tabular}{|l|r|r|r|}
\hline
 & \multicolumn{1}{l|}{\textbf{All}} & \multicolumn{1}{l|}{\textbf{Experts}} & \multicolumn{1}{l|}{\textbf{Non-experts}} \\ \hline
\multicolumn{4}{|c|}{\textbf{Percieved quality of VR experience}} \\ \hline
Avg            & 8.67           & 8.22           & 9.11           \\ \hline
StDev          & 1.029          & 1.202          & 0.601          \\ \hline
CI\protect\footnotemark         & 0.512          & 0.924          & 0.462          \\ \hline
\multicolumn{4}{|c|}{\textbf{Percieved educational value}}        \\ \hline
Avg            & 8.78           & 8.67           & 8.89           \\ \hline
StDev          & 8.88           & 1.00           & 0.78           \\ \hline
CI             & 0.43           & 0.769          & 0.601          \\ \hline
\multicolumn{4}{|c|}{\textbf{Recall activity score}}              \\ \hline
Avg            & 9.26           & 8.93           & 9.63           \\ \hline
StDev          & 1.30           & 1.66           & 0.73           \\ \hline
CI             & 0.65           & 1.28           & 0.56           \\ \hline
\end{tabular}
\caption{VR training application - rating scores}
\end{center}
\end{table}

Time on task and the number of help requests were recorded to support further analysis of participants’ ratings with regard to the effort they invested. All participants were able to complete the steps of the training scenario successfully and within a reasonable time (2:50 minutes on average). Slightly higher completion time was recorded on average for the non-experts, but with no statistical significance (t(8)=-1.5, p=0.17). However, there was a statistically significant difference in the number of help requests between the two groups (t(8)=-4.6, p=0.001), as expected, due to the inexperience of the users.

\begin{table}[!http]
\begin{center}
\begin{tabular}{|l|r|r|r|}
\hline
 & \multicolumn{1}{l|}{\textbf{All}} & \multicolumn{1}{l|}{\textbf{Experts}} & \multicolumn{1}{l|}{\textbf{Non-experts}} \\ \hline
\multicolumn{4}{|c|}{\textbf{Time (min)}}          \\ \hline
Avg          & 2:50       & 2:38       & 3:03      \\ \hline
StDev        & 0:59       & 1:00       & 1:00      \\ \hline
CI           & 0:29       & 0:46       & 0:46      \\ \hline
\multicolumn{4}{|c|}{\textbf{\# of help requests}} \\ \hline
Avg          & 2.56       & 1.78       & 3.33      \\ \hline
StDev        & 1.20       & 0.97       & 0.87      \\ \hline
CI           & 0.60       & 0.75       & 0.67      \\ \hline
\end{tabular}
\caption{VR training application - time on task and number of help requests}
\end{center}
\end{table}
\addtocounter{footnote}{0}
\footnotetext{CI: 95\% Confidence Interval}

\subsubsection{Second session: Visual Scripting tool}
All participants rated highly the perceived easiness of completing the two given script tasks with the Visual Scripting tool. However, paired t-testing revealed a statistically significant difference in the scores received for Task 1 by the SD experts and by the non-experts; t(4)=-5.66, p = 0.005. Similarly, there was a statistically significant difference in the scores received for Task 2 by the SD experts and the non-experts; t(8)=2.8, p=0.02. 

\begin{table}[!http]
\begin{center}
\begin{tabular}{|l|c|c|c|c|c|c|}
\hline
 & \multicolumn{2}{c|}{\textbf{All}} & \multicolumn{2}{c|}{\textbf{Experts}} & \multicolumn{2}{c|}{\textbf{Non-experts}} \\ \hline
\multicolumn{1}{|c|}{} & T1   & T2   & T1   & T2   & T1   & T2   \\ \hline
Avg                    & 8.00 & 8.28 & 8.44 & 8.67 & 7.56 & 7.89 \\ \hline
StDev                  & 1.19 & 0.95 & 1.13 & 1.00 & 1.13 & 0.78 \\ \hline
CI                     & 0.59 & 0.48 & 0.87 & 0.77 & 0.87 & 0.60 \\ \hline
\end{tabular}
\caption{Visual Scripting – perceived task easiness for Task 1 (T1) and Task 2 (T2)}
\end{center}
\end{table}

These observations are aligned with the differences in the measurements of time on task and number of help requests between the SD experts and the non-experts. As shown in Table 4 non-experts required both more time and assistance. Paired t-testing confirmed that the differences carried a statistical significance both for time on task (t(8)=-14.69, p=0.0000004) and number of help requests (t(8)=-2.25, p=0.05). 

Nevertheless, it is interesting that the additional required effort by non-experts to complete the tasks, did not affect their overall experience with the tool. In fact, all participants rated highly the overall experience of using this tool (8.61/10), without any statistical difference between the two groups; t(8)=-1.1, p=0.3.

\begin{table}[!http]
\begin{center}
\begin{tabular}{|l|r|r|r|}
\hline
 & \multicolumn{1}{l|}{\textbf{All}} & \multicolumn{1}{l|}{\textbf{Experts}} & \multicolumn{1}{l|}{\textbf{Non-experts}} \\ \hline
\multicolumn{4}{|c|}{\textbf{Time (min)}}          \\ \hline
Avg          & 12:36      & 9:41      & 17:31      \\ \hline
StDev        & 4:14       & 1:02      & 1:32       \\ \hline
CI           & 2:06       & 0:47      & 1:11       \\ \hline
\multicolumn{4}{|c|}{\textbf{\# of help requests}} \\ \hline
Avg          & 2.89       & 2.11      & 3.67       \\ \hline
StDev        & 1.41       & 1.54      & 0.71       \\ \hline
CI           & 0.70       & 1.18      & 0.54       \\ \hline
\multicolumn{4}{|c|}{\textbf{Overall experience}}  \\ \hline
Avg          & 8.61       & 7.88      & 8.44       \\ \hline
StDev        & 0.98       & 0.78      & 1.13       \\ \hline
CI           & 0.49       & 1.13      & 0.86       \\ \hline
\end{tabular}
\caption{Visual Scripting - time on task, number of help requests (for the entire scenario), and overall experience}
\end{center}
\end{table}

\subsubsection{Third session: VR Editor tool }
The participants also rated highly the perceived easiness of completing the two tasks for the VR Editor tool evaluation. Just like in the Visual Scripting tool, the SD experts found the tasks easier than non-experts, a difference which was identified as statistically important both for Task 1 (t(8)=2.56, p=0.02) and Task 2 (t(8)=2.34, p=0.04).

\begin{table}[!http]
\begin{center}
\begin{tabular}{|l|c|c|c|c|c|c|}
\hline
 & \multicolumn{2}{c|}{\textbf{All}} & \multicolumn{2}{c|}{\textbf{Experts}} & \multicolumn{2}{c|}{\textbf{Non-experts}} \\ \hline
\multicolumn{1}{|c|}{} & T1   & T2   & T1   & T2   & T1   & T2   \\ \hline
Avg                    & 7.50 & 7.06 & 8.29 & 7.67 & 6.89 & 6.44 \\ \hline
StDev                  & 1.26 & 1.39 & 1.38 & 1.66 & 0.78 & 0.73 \\ \hline
CI                     & 0.67 & 0.69 & 1.28 & 1.27 & 0.60 & 0.56 \\ \hline
\end{tabular}
\caption{: VR Editor – perceived task easiness for Task 1 (T1) and Task 2 (T2)}
\end{center}
\end{table}

Differences were exhibited on the time on task and the number of help requests between the SD experts and the non-experts as expected and in alignment with the perceived ease of completing the tasks. Paired t-testing confirmed that the differences carried a statistical significance both for time on task (t(8)=-7.1, p=0.0009) and for the number of help requests: (t(8)=-3.05, p=0.01).

Just like the observed results in the second session with regard to the overall experience, non-experts gave equally high rates to the overall experience as the experts, despite the extra time and effort required to complete the tasks. The overall experience score was 7.61/10, while no statistically important difference between the groups was observed (t(8)=0.5, p=0.6).

\begin{table}[!http]
\begin{center}
\begin{tabular}{|l|r|r|r|}
\hline
 & \multicolumn{1}{l|}{\textbf{All}} & \multicolumn{1}{l|}{\textbf{Experts}} & \multicolumn{1}{l|}{\textbf{Non-experts}} \\ \hline
\multicolumn{4}{|c|}{\textbf{Time (min)}}          \\ \hline
Avg          & 13:24      & 9:08      & 17:39      \\ \hline
StDev        & 5:00       & 2:52      & 2:03       \\ \hline
CI           & 2:29       & 2:12      & 1:35       \\ \hline
\multicolumn{4}{|c|}{\textbf{\# of help requests}} \\ \hline
Avg          & 2.83       & 2.22      & 3.44       \\ \hline
StDev        & 1.04       & 0.97      & 0.73       \\ \hline
CI           & 0.52       & 0.75      & 0.56       \\ \hline
\multicolumn{4}{|c|}{\textbf{Overall experience}}  \\ \hline
Avg          & 7.61       & 7.77      & 7.44       \\ \hline
StDev        & 1.09       & 1.09      & 1.13       \\ \hline
CI           & 0.54       & 0.84      & 0.86       \\ \hline
\end{tabular}
\caption{VR Editor - time on task, number of help requests (for the entire scenario), and overall experience}
\end{center}
\end{table}

\begin{figure}
\begin{center}
  \includegraphics[width=\linewidth]{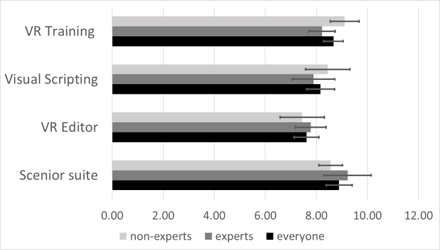}
  \caption{Experience scores for each one of the authoring tools and the overall suite}
  \label{fig:ExperienceScore}
  \end{center}
\end{figure}

In conclusion, all participants regardless of their expertise in VR and SD were able to successfully complete the tasks. The non-experts did – as expected – require more time and assistance, but did not seem to affect their overall experience in using the tools.  The results of the evaluation matched the general sentiment of the participants about the overall suite expressed through the positive comments in semi-structured interviews, as well as through responding to a corresponding question in the questionnaire (Figure~\ref{fig:ExperienceScore}). 

\section{Conclusions and Future Work}
In this work, we presented a novel system capable to generate gamified training experiences exploiting its modular architecture and the authoring tools we developed. We introduced the scenegraph as a dynamic, acyclic data structure to represent any training scenario following an educational curriculum. In addition, we proposed a category of new software design patterns, the Action prototypes, specially formulated for interactive VR applications. Finally, we developed a visual scripting tool along with a VR Editor to enhance the visualization and speed up content creation.

Our system has certain limitations linked with its components and functionalities. First of all, the evaluation process highlighted weaknesses in the interaction with the VR editor. Although it behaves well in Action customization, the script generation process is still complex due to the amount of information and steps needed from the user. In addition, some of its interactive components are not intuitive, resulting in the frustration of users when asked to implement certain behaviors. Finally,  regarding the visual scripting editor, the real-time compilation process may cause performance issues in complex training scenarios and delay the initialization of scenegraph.

The purpose of the evaluation was to get an overall impression of the authoring tools. This did not allow for in-depth analysis of each tool separately, in terms of effectiveness and efficiency. This is a known limitation that will be rectified by conducting further testing for each tool separately in future iterations.

In the future, we aim to utilize computer vision to capture the trainer's movements from external cameras or directly from within the virtual environment to automatically generate interactive behaviors in VR. Another idea is to collect this data through video from a real-life scenario by monitoring the trainer and afterward processing the data using machine learning to extract important key features and construct a template of the training scenario. 

\section*{Acknowledgements}
This project has received funding from the European Union’s Horizon 2020 research and innovation programme under grant agreement No 871793 (ACCORDION) and No727585 (STARS-PCP) and supported by Greek national funds (projects VRADA and vipGPU).

\bibliographystyle{spmpsci}
\bibliography{template} 

\begin{thebibliography}{10}
\bibitem{HOOPER2019}
Our clinical trial (citation not provided for the reviewing process)
\bibitem{MAGES2018}
Our {VR} training platform (citation not provided for the reviewing process)
\end{thebibliography}

\end{document}